\pdfoutput=1 
\documentclass{JINST}
\usepackage{cite}
\usepackage{comment}
\usepackage{epstopdf}
\usepackage{amsmath}
\usepackage{float}
\title{Estimation of low energy neutron flux ($E_n\leq15$ MeV) in India-based Neutrino Observatory cavern using Monte Carlo techniques}

\author{N. Dokania$^{a,b}$, V. Singh$^{a,b}$, S.~Mathimalar$^{a,b}$, A. Garai$^{a,b}$,
V. Nanal$^c$\thanks{Corresponding author.}, R.~G.~Pillay$^c$ and K.~G.~Bhushan$^d$\\
\llap{$^a$}India based Neutrino Observatory,\\
 Tata Institute of Fundamental Research, Mumbai 400 005, India \\
\llap{$^b$}Homi Bhabha National Institute,\\
  Anushaktinagar, Mumbai 400 094, India\\
\llap{$^c$}Department of Nuclear and Atomic Physics, Tata Institute of Fundamental Research,\\
Colaba, Mumbai 400 005, India\\ 
 \llap{$^d$} Technical Physics Division, Bhabha Atomic Research Centre,\\ Mumbai 400 085, India\\

E-mail: \email{nanal@tifr.res.in}}

\abstract{The neutron flux at low energy ($E_n\leq15$ MeV) resulting from the radioactivity of the rock in the underground cavern of the India-based Neutrino Observatory is estimated using Geant4-based Monte Carlo simulations. The neutron production rate due to the spontaneous fission of $^{235, 238}$U, $^{232}$Th and ($\alpha, n$) interactions in the rock is determined employing the actual rock composition. It is shown that the total flux is equivalent to a finite size cylindrical rock ($D=L=140$ cm) element. The energy integrated neutron flux thus obtained at the center of the underground tunnel is 2.76 (0.47) $\times 10^{-6}\rm~n ~cm^{-2}~s^{-1}$. The estimated neutron flux is of the same order ($\sim10^{-6}\rm~n ~cm^{-2}~s^{-1}$) as measured in other underground laboratories.}


\begin{document}

\section{Introduction}\label{sec:intro}
Neutrons are known to be an important source of background for experiments like direct dark matter searches~\cite{dm}, double beta decay experiments~\cite{cremonesi}, solar neutrino measurements~\cite{solar}, etc. In underground laboratories, neutrons originate from the spontaneous fission of U and Th trace elements in the surrounding rock and the ($\alpha, n$) interactions on the low Z elements present in the rock. Neutrons induced by cosmic-ray muon interactions with rock and shielding material are generally more penetrating because of higher energy ($E_n > 15$ MeV) but expected flux is $\sim$ 100--1000 times lower~\cite{bellini1, mei}. To reach the desired sensitivity in the experiment, the neutron background from rock, detector components and cosmic-ray muons should be significantly suppressed. 
The neutron background from rock ($E_n \leq 15$ MeV) can be reduced by installing passive shield of hydrocarbon material or water surrounding the detector. The size and composition of the shield  depends on the low energy neutron flux.

In India, a Tin cryogenic
bolometer (TIN.TIN) is under development to search for neutrinoless double beta decay ($0\nu\beta\beta$) in $^{124}$Sn and is proposed to be installed at the upcoming India-based Neutrino Observatory (INO)~\cite{epj_nanal, ical_white}. The neutron-induced background ($E_n \leq 18$ MeV) in the detector and surrounding materials has been studied using neutron activation techniques~\cite{jinst_neha}. To evaluate the background levels induced by fast neutrons in the underground cavern  and the necessary neutron shield requirements, it is essential to have an understanding of the neutron flux in the underground cavern. 
This work describes the estimation of the neutron flux in the cavern from the rock activity (spontaneous fission and ($\alpha, n$) interactions). A volume source distributed uniformly in a finite size rock element is considered in the Monte Carlo (MC) simulations, where the strength of the source is derived from the INO cavern rock composition. The total neutron flux is then estimated at the center of a 12 m long cylindrical tunnel of 2 m radius. Ref. 7 gives the conceptual layout of the proposed underground caverns.

\section{Neutron Yield in the INO Cavern }\label{sec:expt}

The INO cavern will be located in Bodi West Hills (BWH), Madurai, India~\cite{ino}.
The BWH rock is mainly Charnockite, which is the hardest known rock, having a density of $\sim$2.89 $\rm g/cm^3$. It is essential to know the composition of the rock, particularly the content of $^{235, 238}$U, $^{232}$Th and low Z isotopes to determine the neutron production rate. 
The BWH rock composition is obtained from TOF-SIMS (Time of Flight - Secondary Ion Mass Spectrometry) method~\cite{wilson}. Table~\ref{tab:sims} shows various constituent elements of the BWH rock with respective concentrations.  
\begin{table}[!h]
\begin{center}
\caption[Elemental distributions of BWH rock obtained with TOF-SIMS method]{\label{tab:sims} Elemental distributions of BWH rock obtained with TOF-SIMS method.}
\begin{tabular}{ c | c | c | c }
\hline\hline
Element&   Concentration  & Element & Concentration \\ 
       &  (\% Weight)  &  &  (\% Weight) \\ [0.5ex]\hline

$ ^{12} $C & 2& $ ^{39,40,41} $K & 4\rule{0pt}{3ex}\\
$ ^{23} $Na & 5  & $ ^{40} $Ca & 9.99 \\
 $ ^{24,25,26} $Mg &7  &$ ^{52} $Cr & 0.49 \\
$ ^{27} $Al& 25 &$ ^{56} $Fe  & 1 \\ 
$ ^{28,29,30} $Si  & 40  &$ ^{58,60} $Ni & 0.55 \\  
$ ^{31} $P   &  3 &  $ ^{63,65} $Cu & 1.2 \\ 

 &  &$ ^{120} $Sn & 0.6  \\[0.5ex]  \hline\hline
 \end{tabular}
 \end{center}
\end{table}
Since the SIMS method has limited sensitivity $\geq100$ ppb, the Inductively Coupled Plasma Mass Spectrometry (ICPMS) method~\cite{icpms} was used to reliably extract the  $^{235, 238}$U, $^{232}$Th concentration in the rock. It was found that the BWH rock contains 60~ppb of $^{238}$U and 224~ppb of $^{232}$Th. Errors from SIMS and ICPMS measurements are about 1$\%$. In the present work, the BWH rock is assumed to be a homogeneous mixture of its constituent elements and the $^{235, 238}$U, $^{232}$Th impurities are distributed uniformly in the rock.
Due to very small isotopic abundance of $^{235}$U (0.72\%) and low SF probability ($\sim7\times10^{-9}$~\cite{nndc}), its contribution to the neutron flux is expected to be negligible.
Hence, only $^{238}$U and $^{232}$Th decays are considered here.
The spectrum of the neutrons emitted in Spontaneous fission (SF) is described by an analytic function, known as Watt spectrum~\cite{watt} and is given by:
\begin{align}\label{eqn:watt}
W(a,b,E) &= C e^{-E/a} sinh (\sqrt{b E}) \\
{\rm where}~~C &= {\sqrt(\frac{{\pi} b}{4 a}}) (\frac{e^{b/4a}}{a})
\end{align}
The parameters $a$ and $b$ are empirically derived for each isotope~\cite{jerome} (see Table~\ref{tab:wattab}). 
\begin{table}[!h]
\begin{center}
\caption[The Watt spectrum parameters for $^{238}$U and $^{232}$Th]{\label{tab:wattab} The Watt spectrum parameters for $^{238}$U and $^{232}$Th~\cite{jerome}.}
\begin{tabular}{  c | c |c}
\hline\hline
Isotope &  a & b \\ 
       &  (MeV) & (MeV$^{-1}$)\\ [0.5ex]\hline
$^{238}$U & 0.64832 & 6.81057 \rule{0pt}{3ex} \\
$^{232}$Th  & 0.80 & 4.0  \\[0.5ex] \hline\hline
\end{tabular}
 \end{center}
\end{table}

\begin{figure}[!h]
\begin{centering}
\includegraphics[scale=0.6, trim=0.1cm 0cm 0cm 0cm, clip=true]{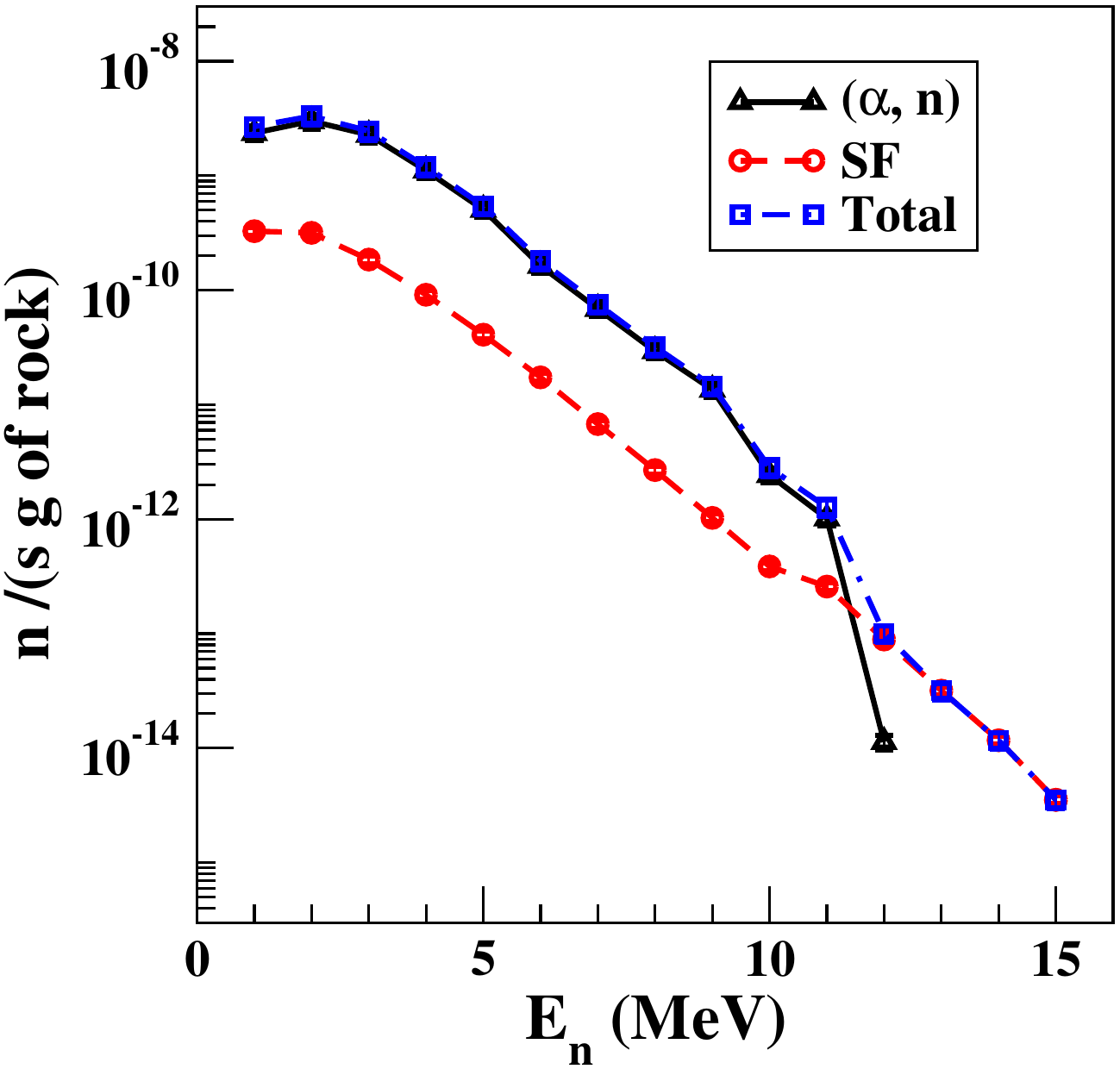} 
\caption{\label{fig:fiss} (Color online) Neutron spectra for BWH rock with 60~ppb of $^{238}$U and 224 ppb of $^{232}$Th. Errors are within the point size.}
\end{centering}
\end{figure}

The ($\alpha, n$) reaction rate depends on the initial energy of the emitted alpha particle, the reaction $Q$ value and the Coulomb barrier. 
For the present work, the thick target ($\alpha$, n) reaction yields ($N(E)$ - neutrons per MeV) have been taken from Refs. \cite{alpha_mei, link}. The total thick target neutron yield is determined by the sum of the individual element yield weighted by its mass ratio in the BWH rock (as per Table~\ref{tab:sims}). The neutron yields thus obtained, normalized to $^{238}$U and  $^{232}$Th content of the BWH rock, are shown in Figure~\ref{fig:fiss}.
It is evident that the ($\alpha, n$) component dominates at lower neutron energies while the SF dominates at higher energies ($E_n > 11$ MeV). The yield for ($\alpha, n$) falls by six orders of magnitude for $E_n > 12$ MeV and can be neglected. The uncertainties assumed in the ($\alpha, n$) and the spontaneous fission yields are 10\% and 8\%, respectively~\cite{watt}.

\section{Monte Carlo Simulations of Neutron Flux}
The underground observatory at the INO site will have a rock cover of at least 1000 m on all sides. However, the neutron flux will get attenuated in few meters of rock~\cite{lemrani} and hence only a finite size rock element will contribute to the neutron flux in the cavern. The size of the rock element is selected in such a way that the volume is sufficient to achieve saturation after scattering from within the rock element. In other words, neutrons originating outside this volume element have a negligible contribution to the flux at surface of the rock element. For simplicity of simulations, a cylindrical geometry is considered. As can be seen from Figure~\ref{fig:fiss}, neutron source in the rock material is very weak and the flux rapidly decreases with energy. This will necessitate the large scale simulations and will be prone to errors due to statistical fluctuations. To overcome this problem, an alternative approach using MC simulations has been employed.
For mono-energetic, uniformly distributed, isotropically emitted neutrons within the volume of the rock element, energy spectra ($N(E_j)$) after transmission through a rock element were computed. Since the neutron flux rapidly decreases with energy (by about $\sim$ six orders of magnitude at $E_n=15$ MeV), only 
$E_n\le$15~MeV is considered in the simulations. The transmission factor $T(E_j, E_i)$ -- the fraction of neutrons with initial energy $E_i$ and emerging with final energy $E_j$ is computed for $E_i=1$ to 15 MeV. The size of the rock element was optimized such that $T(E_j, E_i)$ reaches a saturation value for the highest incident neutron energy $E_n=15$ MeV, over the entire spectrum region ($E_j$). The neutron flux $N(E_j)$ (in units of $\rm~g^{-1}~s^{-1}$) originating from the optimized rock element is then computed using:

\begin{equation}
\label{eq:matrix}
N(E_j) = \sum_{i}{T(E_j, E_i)N(E_i)} 
\end{equation}
where $N(E_i)$ is taken from the total neutron spectrum of the BWH rock given in Figure~\ref{fig:fiss}.
The neutron flux $N_s(E_j)$ (in units of $\rm~cm^{-2}~s^{-1}$) per unit area at the surface of the detector, over which the transmission of neutrons from the rock element is recorded, is given by:

\begin{equation}
\label{eq:density}
N_s(E_j) = \frac{\rho\,V\,N(E_j)}{A}
\end{equation}
where $\rho$ is the density of the INO cavern rock, $V$ is the volume of the optimized rock element and $A$ is the surface area of the detector kept in front of the rock element.
Finally the neutron flux seen by the detector is estimated at the center of the tunnel by integrating over the entire cylindrical surface of the tunnel. 

The GEANT4-based MC simulation studies have been done using the G4NDL4.2 neutron cross-section library~\cite{geant}. Typically, $10^6$ events were generated at each incident energy and final energy spectra is made with bin widths of 1 MeV.

\section{Neutron Transmission through the INO cavern rock}

\begin{figure}[!h]
\begin{centering}
\includegraphics[scale=0.5, trim=0.01cm 0cm 0cm 0cm, clip=true]{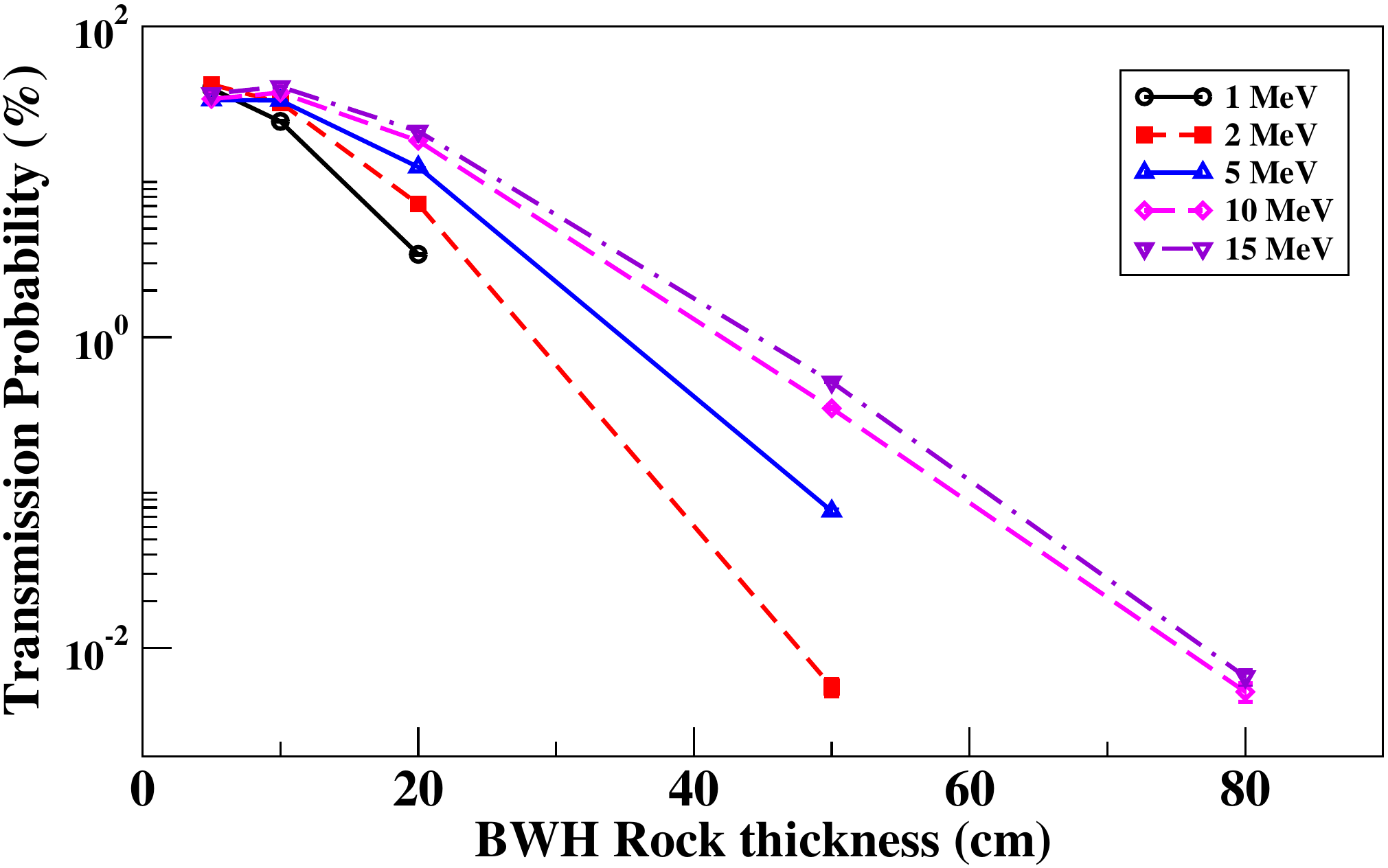} 
\caption{\label{fig:vecc}(Color online) Neutron transmission probability as a function of rock thickness for a mono-energetic point source. Lines are only a guide to the eye. } 
\end{centering}
\end{figure}

\begin{figure}[H]
\begin{centering}
\includegraphics[scale=0.3]{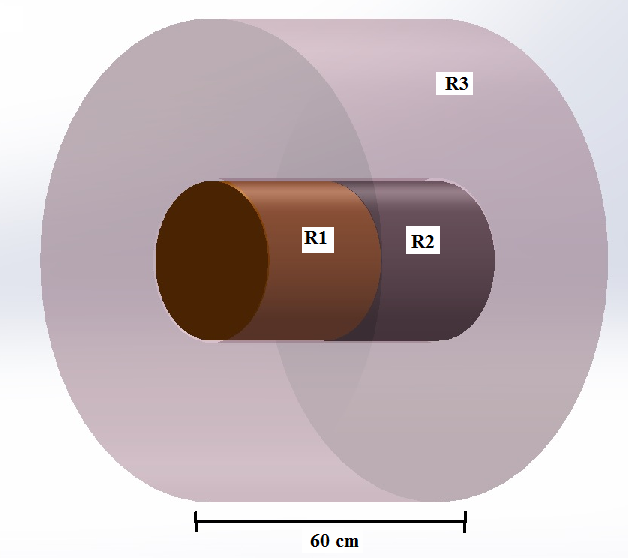} 
\caption{\label{fig:r123} (Color online) Schematic geometry of rock elements $R_1$, $R_2$ and $R_3$ considered in the simulations.}
\end{centering}
\end{figure}

\begin{figure}[!h]
\begin{centering}
\includegraphics[scale=0.5, trim = 0.01cm 0cm 0cm 0cm, clip=true]{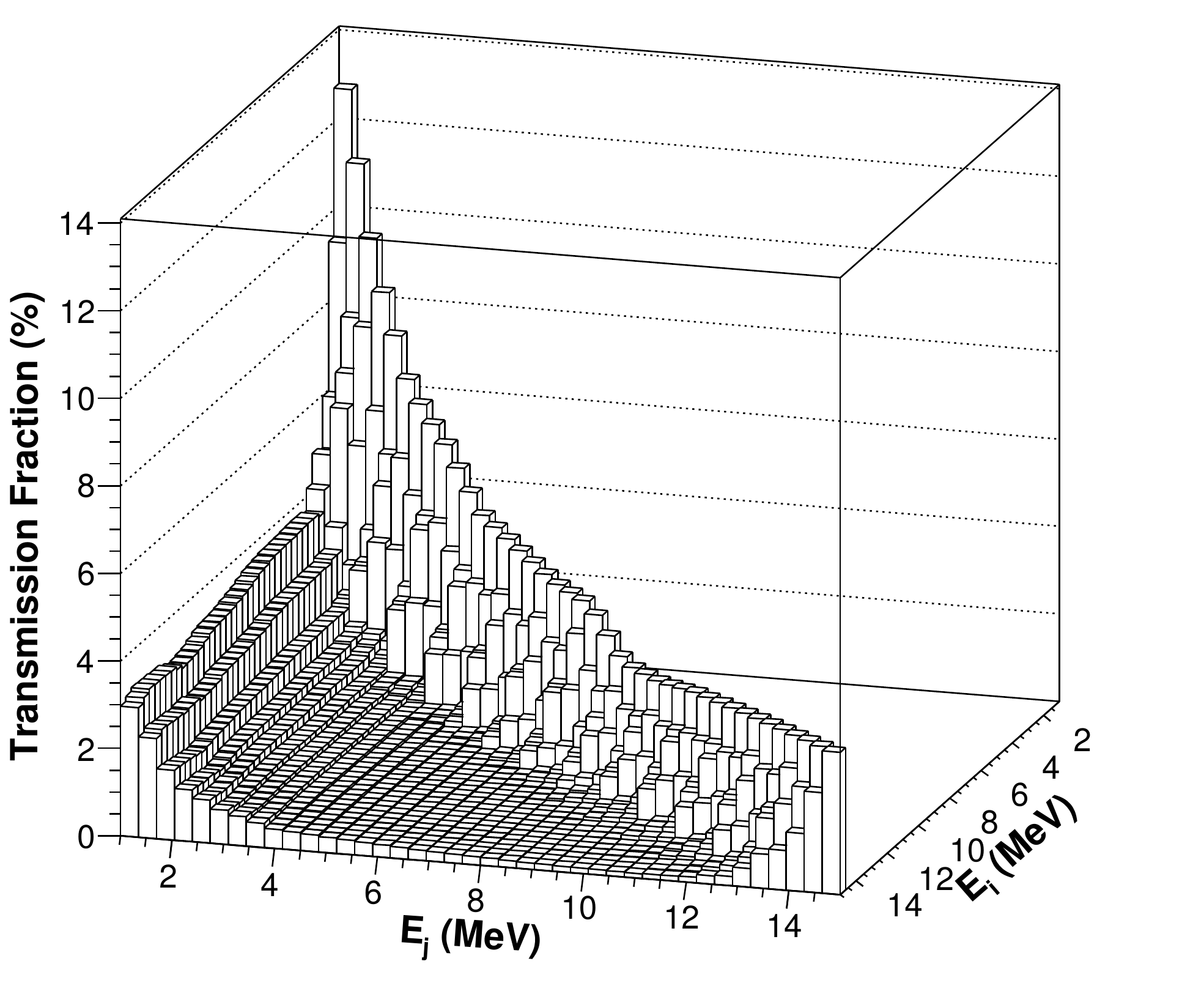} 
\caption{\label{fig:panel_10}Transmission fraction as a function of incident neutron energy $E_i=1-15$ MeV and the corresponding transmitted neutron energy ($E_j$) after propagation through $R_1$ (Case 1).}
\end{centering}
\end{figure}

To understand the effects of neutron attenuation and scattering as a function of neutron energy in the rock, simulations were done for different configurations. For investigating the attenuation effect, a mono-energetic point source of neutrons placed behind the center of $40\times40~\rm cm^2$ BWH rock of different thicknesses was considered.
Figure~\ref{fig:vecc} shows the neutron transmission probability ($P(E_i)$) as a function of rock thickness for $E_n=1-15$ MeV. It can be seen that the flux of neutrons of $E_n=15$ MeV reduces by about two orders of magnitude after propagation through 30 cm thick BWH rock.
It should be pointed out that $P(E_i)$ decreases because neutrons lose energy, and as a result the yield of low energy neutrons ($N(E_j), E_j<E_i$) is enhanced (see Figure~\ref{fig:panel_10}). However, the low energy neutron yield is also affected by scattering in surrounding material.
To understand the effect of the surrounding material, the rock volume was subdivided into inner cylinder and outer cylindrical shell. The inner cylinder was further divided into two cylinders $R_1$ and $R_2$ with diameter $D=30$ cm and length $L=30$ cm. The outer shell ($R_3$) dimensions were chosen to be $D=90\rm~cm \rm~and~L=60$ cm. These dimensions were chosen on the basis of Figure~\ref{fig:vecc}. The schematic geometry of these sub-divided volumes ($R_1$, $R_2$, $R_3$) is shown in Figure~\ref{fig:r123}. A thin detector of size $D=30\rm~cm, L=0.2$ cm is placed on the face of the $R_1$ for recording transmitted neutrons.
Simulations were done for the following three configurations with a uniform volume source : 
\begin{itemize}
\item{Case 1: $R_1$-rock, $R_2$ and $R_3$-Air : Neutrons in $R_1$.}
This configuration studies the neutron propagation through $R_1$.
\item{Case 2: $R_1$ and $R_2$-rock, $R_3$-Air : Neutrons in $R_2$.}
The case shows the attenuation of neutrons after propagation through rock volume $R_1$.
\item{Case 3: $R_1$, $R_2$ and $R_3$-rock : Neutrons in $R_1$.}
The case illustrates the effect of scattering of neutrons from adjacent rock volumes $R_2$ and $R_3$.
\end{itemize}
A typical spectrum for the first configuration is shown in Figure~\ref{fig:panel_10}.

\begin{figure}[H]
        \begin{minipage}[b]{0.45\linewidth}
          \centering
          \includegraphics[scale=0.5, trim=0.01cm 0cm 0cm 0cm, clip=true]{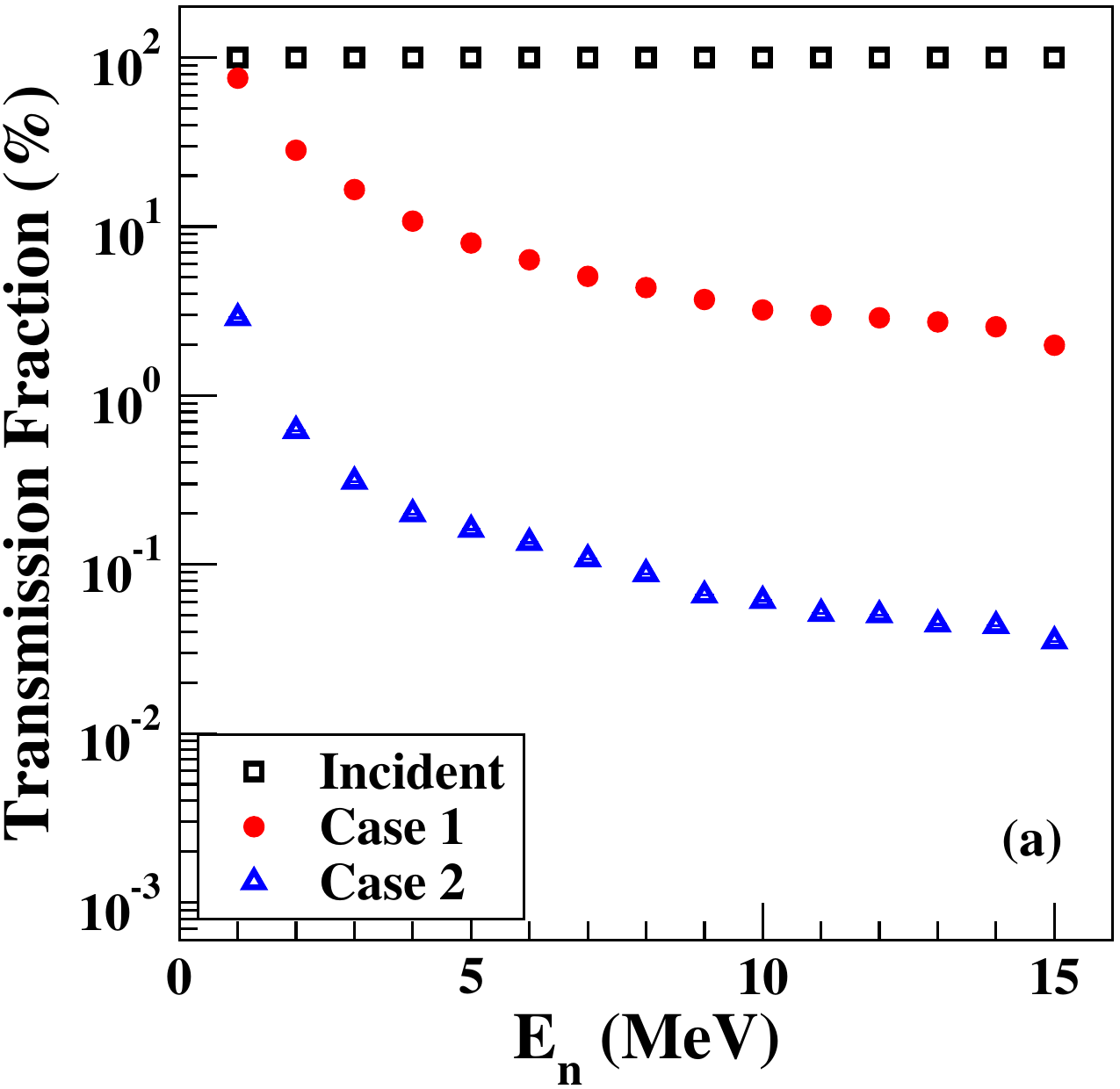}
          \end{minipage}
        \begin{minipage}[b]{0.45\linewidth}
          \centering
          \includegraphics[scale=0.5,  trim=0.01cm 0cm 0cm 0cm, clip=true]{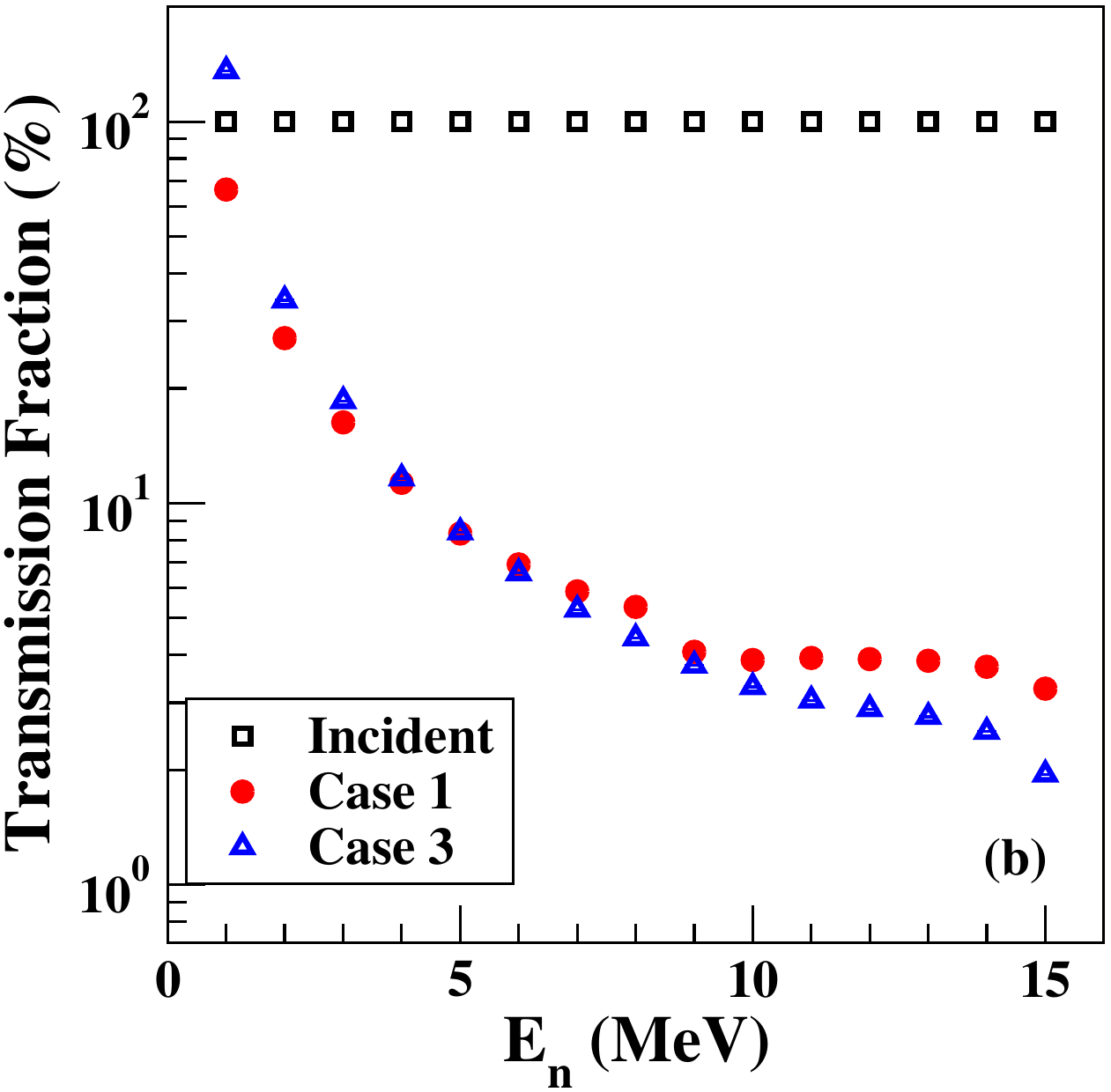}
               \end{minipage}
        \caption{\label{fig:rock_60cm}(Color online) (a) Neutron transmission shown for Case 1 (red circle) and Case 2 (blue triangle) and 
        (b) Neutron transmission shown for Case 1 (red circle), Case 3 (blue triangle) (see text for details)  }
      \end{figure}

The total transmitted neutron spectrum was generated using :
\begin{align}\label{eq:delta}
\nonumber
N(E_j) &=\sum_{i}T(E_j, E_i)\,N (E_i)\\
{\rm where}~~N(E_i)& =N_0\,\delta(E_i)
\end{align}
 The spectra for all three configurations are shown in the Figure~\ref{fig:rock_60cm}.
      
It can be seen from Figure~\ref{fig:rock_60cm}(a) that the overall contribution from the $R_2$ is only about 10\% of that of the $R_1$. Further, reduction in the flux at higher energies ($E_n>10$ MeV) is greater than that at lower energies. The enhancement in the low energy yield arising due to scattering in surrounding rock material is clearly visible in Figure~\ref{fig:rock_60cm}(b) in the spectrum corresponding to case 3. 
It is expected that after certain thickness the contribution due to scattering effects will saturate which is demonstrated in the next section.

\section{Rock Element Size Optimization}
\begin{figure}[!h]
\begin{centering}
\includegraphics[scale=0.55, trim = 0.01cm 0cm 0cm 0cm, clip=true]{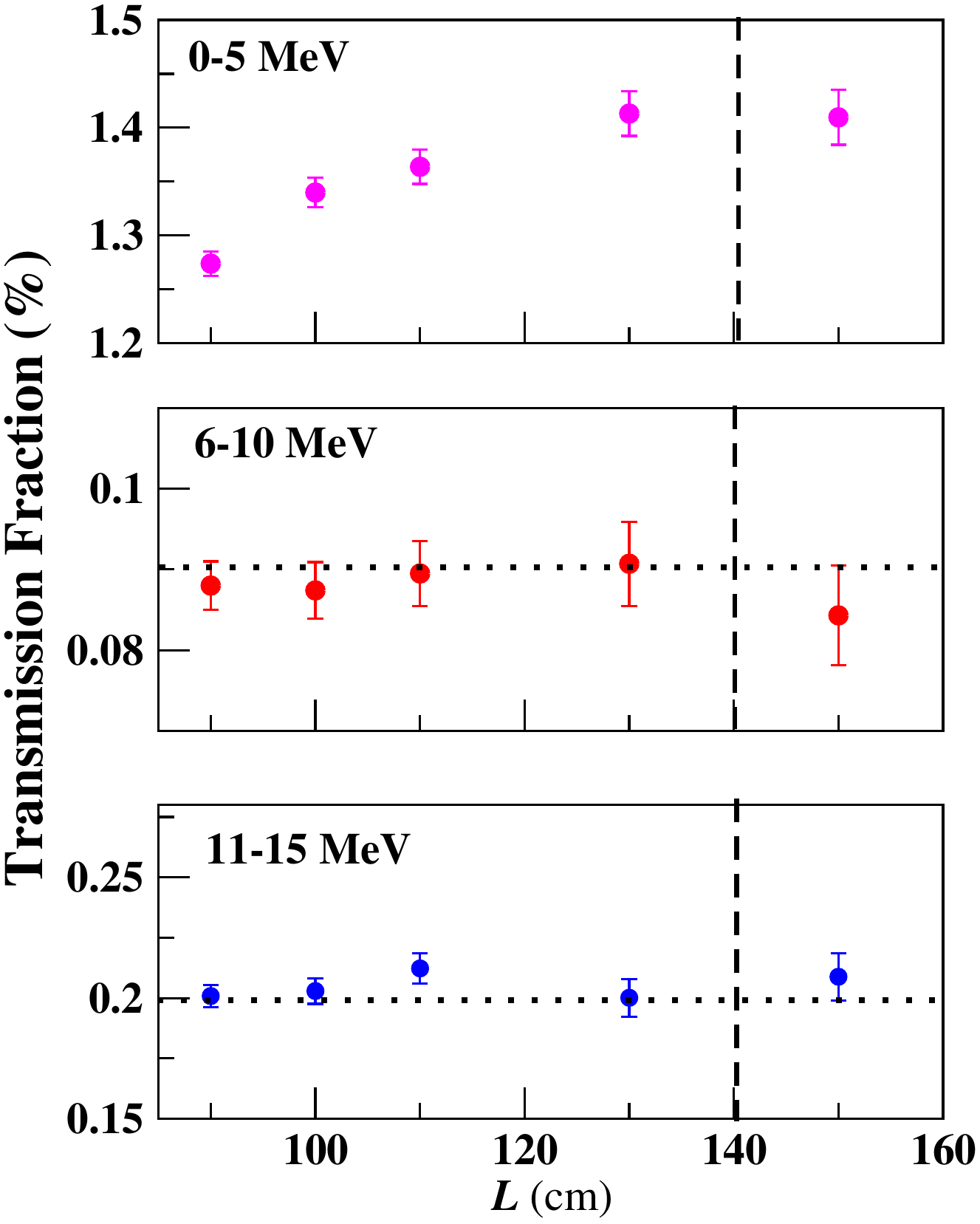} 
\caption{\label{fig:finite} (Color online) Transmission fraction of neutron for $E_n=15$ MeV as a function of $D=L$ of the rock element for the different energy ranges. Each rock element has been scaled to the volume of an element with $D=L=$ 90 cm.}
\end{centering}
\end{figure}

In order to find the rock thickness where the scattering effects are saturated, simulations were done for cylindrical rock elements of different diameters ($D$ - 90, 100, 110, 130, 150 cm). In each case, the length of the cylinder was kept same as the diameter to ensure equal transmission length in all directions. Since the scattering is expected to enhance the low energy yield, the transmission fractions for different energy ranges (0-5, 6-10, 11-15 MeV) were studied for each geometry and is shown in Figure~\ref{fig:finite}.

\begin{figure}[!h]
\begin{centering}
\includegraphics[scale=0.5, trim = 0.01cm 0cm 0cm 0cm, clip=true]{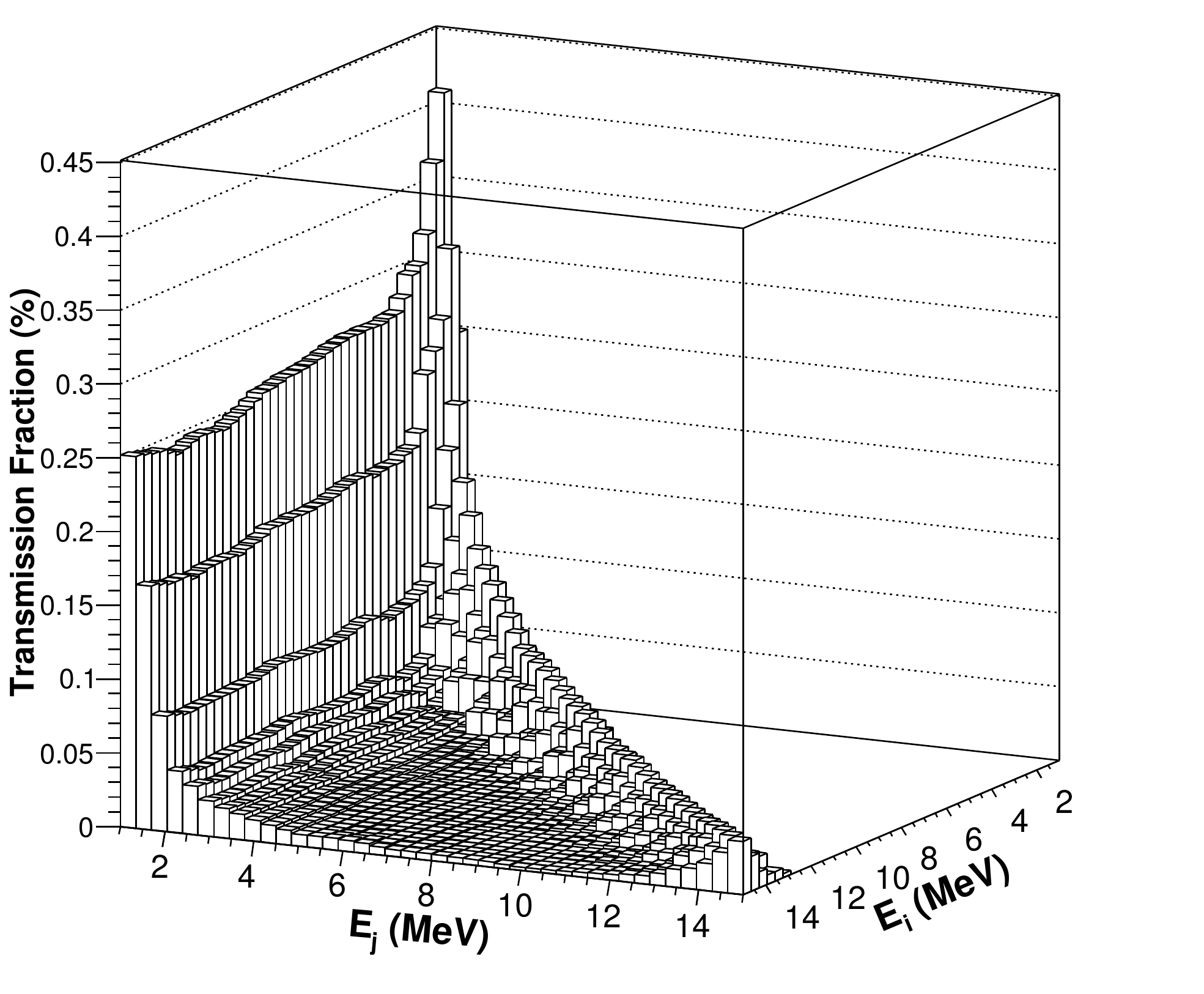} 
\caption{\label{fig:bigpanel}Transmission fraction as a function of incident neutron energy $E_i=1-15$ MeV and the corresponding transmitted neutron energy ($E_j$) for rock element $D = L =$
140 cm.}
\end{centering}
\end{figure}

\begin{figure}[!h]
\begin{center}
\includegraphics[scale=0.43, trim=0.cm 0cm 0cm 0cm, clip=true]{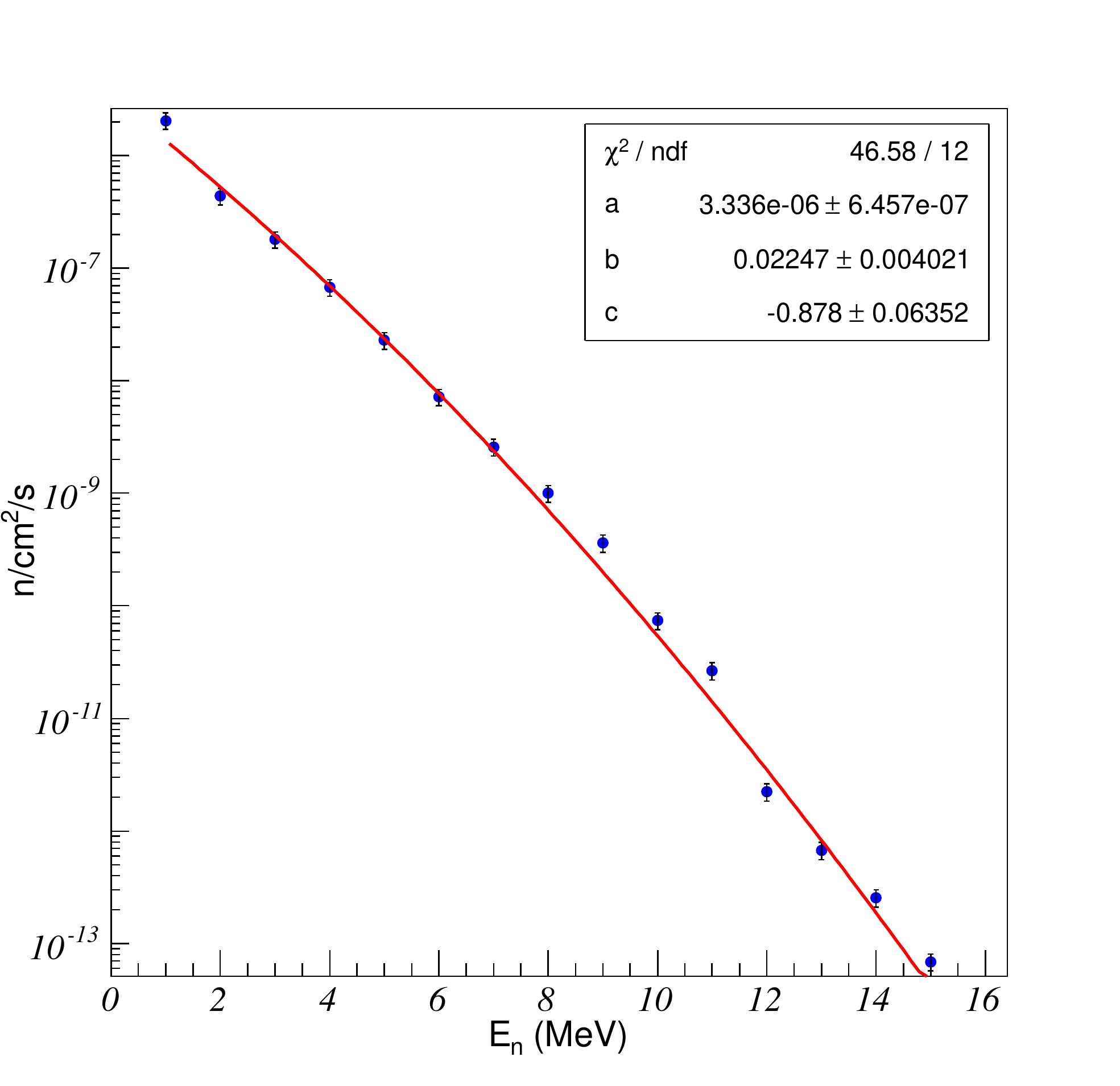} 
\caption{\label{fig:spectra}(Color online) The estimated neutron flux ($N_f(E)$) at the center of an underground tunnel at INO site. The blue points represent the MC data and the red line corresponds to the fit (see text for details).}
\end{center}
\end{figure}

While the transmission fraction is nearly constant for high neutron energies ($E_n>5$ MeV), the saturation value in the low energies window ($E_n=1-5$ MeV) is reached only for rock elements size $D=L>$ 130 cm. Hence the optimal size of rock element is chosen to be a cylinder with $D=L=$ 140 cm. Figure~\ref{fig:bigpanel} shows the neutron spectra from this unit rock element used to generate transmission matrix $T(E_j, E_i)$ as mentioned earlier.
The total neutron flux $N_s (E_j)$ is obtained at the surface of this BWH unit rock element using Eq.~\ref{eq:matrix}. 
It should be noted that the $N_s(E)$ thus obtained represents the flux at any point on the surface of the cylindrical tunnel.
\begin{figure}[H]
\begin{center}
\includegraphics[scale=0.3, trim=0.1cm 0cm 0cm 0cm, clip=true]{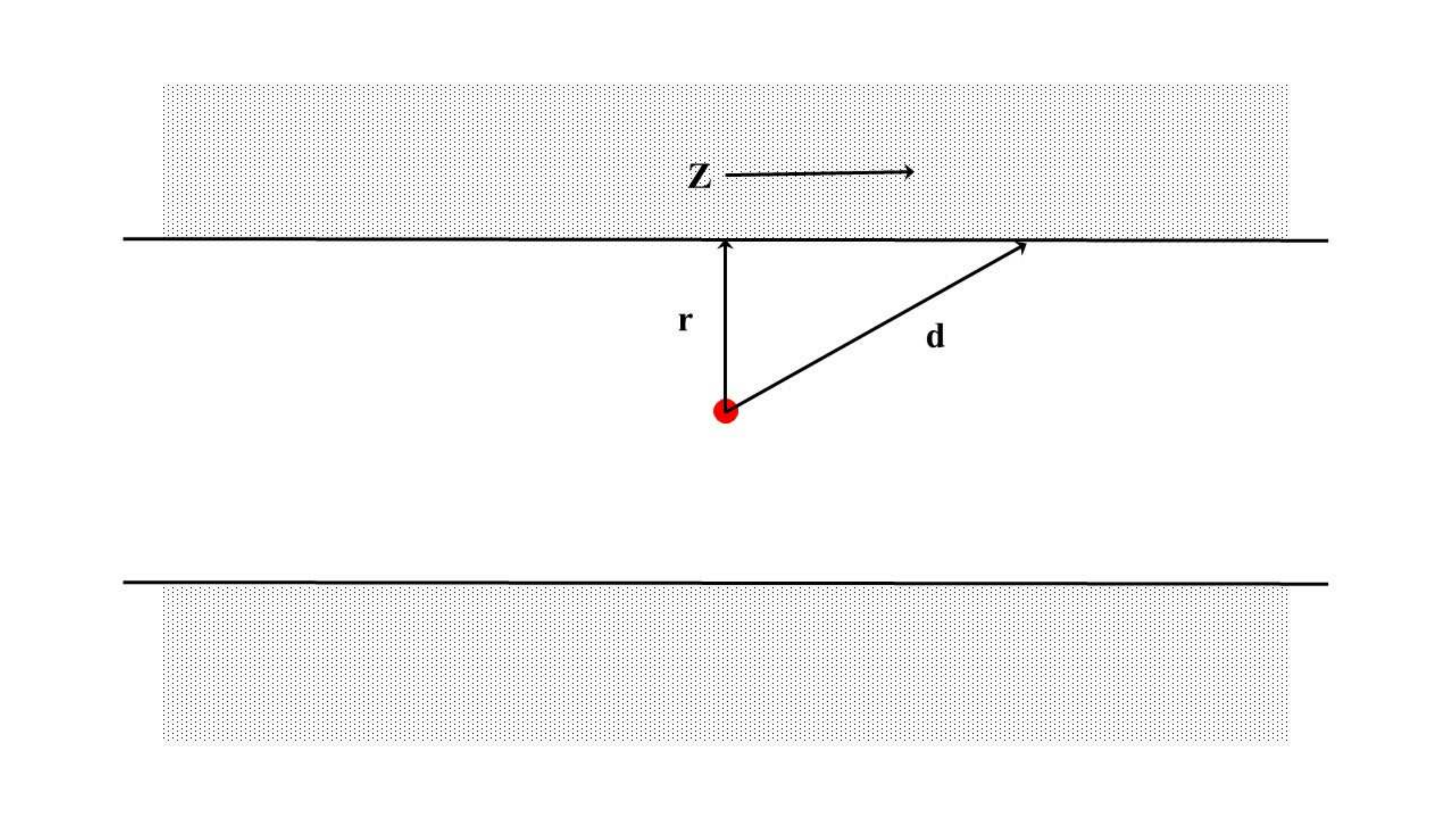} 
\caption{\label{fig:cylinder} (Color online) A schematic diagram of the underground tunnel assumed as a cylinder. The surrounding rock is shown by the shaded portion.}
\end{center}
\end{figure}
Assuming the experimental hall to be a cylindrical tunnel of radius $r=2$ m and $L=12$ m (see Figure~\ref{fig:cylinder}), the neutron flux $N_f(E)$ ($cm^{-2}s{-1}$) per unit area at its center can be estimated as :

\newcommand{\Int}{\int\limits}
\label{eqn:flux_calculation}
\begin{align}
N_f(E)&=\Int_{\phi =0}^{ 2\pi}\Int_{z=-6}^{6}{N_s(E)}~\frac{r\,d\phi\,dz}{d^2} \\
&=N_s(E)\,2\,\pi\Int_{z=-6}^{6}\frac{r}{r^2+z^2}\,dz \\
&=N_s(E)\,2\,\pi~[\arctan{(\frac{z}{r})}]_{z=-6}^{6}\\
&=7.84 \times N_s(E)
\end{align}
The half length of the cylinder was restricted to 6 m since the contribution to the flux from larger lengths will be smaller than 10\%. Moreover, additional shield could be provided at this distance in the tunnel.
Figure~\ref{fig:spectra} shows the estimated neutron flux $N_f(E)$ at the center of the tunnel in the range $E_n=1-15$ MeV. 
The energy integrated neutron flux obtained is $2.76  ~(0.47)\times10^{-6}\rm~n~cm^{-2}~s^{-1}$ from the rock activity. The errors on the flux include the uncertainties in the MC statistical errors and hence the transmission fractions (5\%), SIMS and ICPMS measurement and the neutron yields. This calculated spectrum can be approximately described by a function: 
\begin{equation}
\label{eq:exp_fit}
N(E) = a~\,exp(-(b\,E-c)\,E)
\end{equation}
It should be mentioned that neutrons will also originate from the radioactivity of the concrete of the experimental hall, the exact composition of which is presently not known. The variation of composition of the rock, if any, is also not considered in the present work. The neutron fluxes from the rock measured in other underground laboratories are $3.78  \times10^{-6}\rm~n~cm^{-2}~s^{-1}$ in the Gran Sasso laboratory~\cite{watt} and $1.72\pm 0.61~(\rm stat.)\pm 0.38~(\rm syst.) \times10^{-6}\rm~n~cm^{-2}~s^{-1}$ in the Boulby underground laboratory~\cite{boulby} .

\section{Conclusions}
The neutron flux ($E_n\leq15$ MeV) resulting from spontaneous fission and ($\alpha, n$) interactions for BWH rock containing 60 ppb of $^{238}$U and 224 ppb of $^{232}$Th is estimated. Since the neutron source in the rock material is very weak and the flux rapidly decreases with energy, an alternative approach using MC simulations has been employed. It is shown that only finite rock size contributes to the neutron flux at surface and a rock element of size $D=L=140$ cm is optimal to evaluate the neutron flux. The total neutron flux at the center of a 2 m radius, 12 m long tunnel in the underground cavern is determined. The estimated flux at low energy ($E_n\leq15$ MeV) is 2.76 (0.47) $\times 10^{-6}\rm~n ~cm^{-2}~s^{-1}$.  The estimated neutron flux is of the same order ($\sim10^{-6}\rm~n ~cm^{-2}~s^{-1}$) as measured in other underground laboratories. This will help in the study of neutron-induced background and design of neutron shield for TIN.TIN detector. The work will be further extended to study the muon-induced neutron flux in the INO cavern. In addition, experiments will be carried out to measure the neutron flux when the cavern becomes accessible in future.

\acknowledgments
We are grateful to the INO and TIN.TIN collaborations.

\end{document}